\documentclass[12pt]{article}

 \usepackage{graphicx}
 \usepackage{cite}
 \usepackage{amssymb}   %%% note: relevant for "mathbb" and alike
 \usepackage{booktabs}  %%% note: for top/bottom/mid-rule instead of hline in tables
 \usepackage{orcidlink} %%% note: contains "hyperref" as subset

\begin{document}

%%%%%%%%%%%%%%%%%%%%%%%%%%%%%%%%%%%%%%%%%%%%%%%%%%%%%%%%%%%%%%%%%%%%%%%%%%%%%%%

\begin{center}%
{\LARGE\bf Topological susceptibility and\\[2mm] excess kurtosis in SU(3) Yang-Mills theory}
\end{center}%

\vspace{10pt} %%%%%%%%%%

\begin{center}%
{\large\bf Stephan D\"urr\,\orcidlink{0000-0001-5168-5669}\,${}^{a,b}$} {\large and}
{\large\bf Gianluca  Fuwa\,\orcidlink{0000-0002-8195-4900}\,${}^a$}
\\[10pt]
${}^a${\sl Department of Physics, University of Wuppertal, 42119 Wuppertal, Germany}\\
${}^b${\sl J\"ulich Supercomputing Centre, Forschungszentrum J\"ulich, 52425 J\"ulich, Germany}
\end{center}%

\vspace{00pt} %%%%%%%%%%

\begin{abstract}
\noindent
We present a high-precision study of the topological susceptibility in $SU(3)$ pure gauge theory in four space-time dimensions.
The result is based on ensembles at seven lattice spacings and in seven physical volumes to facilitate a controlled continuum and infinite-volume extrapolation.
We use a gluonic topological charge measurement, with gradient flow smoothing in the operator.
Two complementary smoothing strategies are used (one keeps the flow time fixed in lattice units, one in physical units).
Our data support the idea that both strategies yield a universal continuum limit;
we find $\chi_\mathrm{top}^{1/4}r_0=0.4775(14)(11)$ or $\chi_\mathrm{top}^{1/4}=198.1(0.7)(2.7)\,\mathrm{MeV}$.
Our appendix data suggest that the excess kurtosis $\langle q^4 \rangle / \langle q^2 \rangle^2-3$ decreases $\propto L^{-2}$ for large box sizes $L$.
\end{abstract}

%%% v2: some runs (in particular central ensemble L/a=18,beta=6.1912,7_stout) prolonged, now correcting for (tiny) finite-volume effects, text improved.
%%% v2: primary data available at github.com/GianlucaFuwa/topology_data_25
\vspace{00pt} %%%%%%%%%%

\newcommand{\pad}{\partial}
\newcommand{\hqu}{\hbar}
\newcommand{\til}{\tilde}
\newcommand{\pri}{^\prime}
\renewcommand{\dag}{^\dagger}
\newcommand{\<}{\langle}
\renewcommand{\>}{\rangle}
\newcommand{\gaf}{\gamma_5}
\newcommand{\nab}{\nabla}
\newcommand{\lap}{\triangle}
\newcommand{\dal}{{\sqcap\!\!\!\!\sqcup}}
\newcommand{\trc}{\mathrm{tr}}
\newcommand{\Trc}{\mathrm{Tr}}
\newcommand{\Mpi}{M_\pi}
\newcommand{\Fpi}{F_\pi}
\newcommand{\Mka}{M_K}
\newcommand{\Fka}{F_K}
\newcommand{\Met}{M_\et}
\newcommand{\Fet}{F_\et}
\newcommand{\Mss}{M_{\bar{s}s}}
\newcommand{\Fss}{F_{\bar{s}s}}
\newcommand{\Mcc}{M_{\bar{c}c}}
\newcommand{\Fcc}{F_{\bar{c}c}}

\newcommand{\al}{\alpha}
\newcommand{\be}{\beta}
\newcommand{\ga}{\gamma}
\newcommand{\de}{\delta}
\newcommand{\ep}{\epsilon}
\newcommand{\ve}{\varepsilon}
\newcommand{\ze}{\zeta}
\newcommand{\et}{\eta}
\renewcommand{\th}{\theta}
\newcommand{\vt}{\vartheta}
\newcommand{\io}{\iota}
\newcommand{\ka}{\kappa}
\newcommand{\la}{\lambda}
\newcommand{\rh}{\rho}
\newcommand{\vr}{\varrho}
\newcommand{\si}{\sigma}
\newcommand{\ta}{\tau}
\newcommand{\ph}{\phi}
\newcommand{\vp}{\varphi}
\newcommand{\ch}{\chi}
\newcommand{\ps}{\psi}
\newcommand{\om}{\omega}

\newcommand{\psb}{\bar{\psi}}
\newcommand{\etb}{\bar{\eta}}
\newcommand{\psd}{\psi^{\dagger}}
\newcommand{\etd}{\eta^{\dagger}}
\newcommand{\qh}{\hat{q}}
\newcommand{\kh}{\hat{k}}

\newcommand{\bdm}{\begin{displaymath}}
\newcommand{\edm}{\end{displaymath}}
\newcommand{\bea}{\begin{eqnarray}}
\newcommand{\eea}{\end{eqnarray}}
\newcommand{\beq}{\begin{equation}}
\newcommand{\eeq}{\end{equation}}

\newcommand{\mr}{\mathrm}
\newcommand{\mb}{\mathbf}
\newcommand{\ri}{\mr{i}}
\newcommand{\rd}{\mr{d}}
\newcommand{\Nf}{{N_{\!f}}}%{{N_{\!f}}}
\newcommand{\Nc}{N_{ c }}%{{N_{ c }}}
\newcommand{\Nv}{N_{ v }}%{N_\mr{vec}}
\newcommand{\Nx}{N_x}
\newcommand{\Ny}{N_y}
\newcommand{\Nz}{N_z}
\newcommand{\Nt}{N_t}
\newcommand{\Nthr}{N_\mr{thr}}
\newcommand{\Nrhs}{N_\mr{rhs}}
\newcommand{\DS}{D_\mr{S}}
\newcommand{\DW}{D_\mr{W}}
\newcommand{\DA}{D_\mr{A}}
\newcommand{\DB}{D_\mr{B}}
\newcommand{\DKW}{D_\mr{KW}}
\newcommand{\DBC}{D_\mr{BC}}
\newcommand{\MeV}{\,\mr{MeV}}
\newcommand{\GeV}{\,\mr{GeV}}
\newcommand{\fm}{\,\mr{fm}}
\newcommand{\MSbar}{\overline{\mr{MS}}}

%\definecolor{Gray}{rgb}{0.5,0.5,0.5} \newcommand{\gry}{\color{Gray}}
%\definecolor{Red}{rgb}{0.9,0.0,0.0}  \newcommand{\red}{\color{Red}}
%\definecolor{mygreen}{rgb}{0,0.6,0}
%\definecolor{mygray}{rgb}{0.5,0.5,0.5}
%\definecolor{mymauve}{rgb}{0.58,0,0.82}
%\definecolor{myblue}{rgb}{0.0,0.0,0.7}

\hyphenation{topo-lo-gi-cal simu-la-tion theo-re-ti-cal mini-mum con-tinu-um}

%%%%%%%%%%%%%%%%%%%%%%%%%%%%%%%%%%%%%%%%%%%%%%%%%%%%%%%%%%%%%%%%%%%%%%%%%%%%%%%

\section{Introduction\label{sec:intro}}

%%%%%%%%%%%%%%%%%%%%%%%%%%%%%%%%%%%%%%%%%%%%%%%%%%%%%%%%%%%%%%%%%%%%%%%%%%%%%%%

Yang-Mills (YM) theories in four space-time dimensions dynamically generate a scale by a process called ``dimensional transmutation'' \cite{Coleman:1985rnk}.
This scale reflects itself in any dimensionful quantity, for instance the topological susceptibility $\chi_\mr{top}=\lim_{V\to\infty}\<q^2\>/V$.
Here $q$ is the (global) topological charge of the gauge background (see below) and $V$ the volume of the four-dimensional Euclidean box.

The topological susceptibility has the dimension $\MeV^4$.
Our goal is to calculate, from first principles, the ratio between $\chi_\mr{top}^{1/4}$ and another dimensionful quantity for $\Nc=3$ colors (the result has no free parameters).
In this work we use the Sommer radius $r_0$ \cite{Sommer:1993ce} to set the scale, since this facilitates comparison with previous works (see below).
In practical terms this means that we shall calculate the dimensionless quantity $\chi_\mr{top}^{1/4} r_0$ for a number of lattice spacings $a/r_0$
(again in dimensionless units) so that we can extrapolate the results with $(a/r_0)^2\to0$ to the continuum.

The motivation to study the topological susceptibility in QCD-like theories is twofold.
On the one hand, $\chi_\mr{top}$ serves as a vacuum diagnostics tool.
In YM theory it depends on $\Nc$, while in QCD it depends on $\Nc$ and the $\Nf$ individual quark masses (see Refs.~\cite{Leutwyler:1992yt,Durr:2001ty} for details).
On the other hand, the YM susceptibility appears in the Witten-Veneziano formula \cite{Witten:1979vv,Veneziano:1979ec}
\beq
\chi_\mr{top}^\mr{YM} \doteq \frac{F^2}{2\Nf}(M_{\et'}^2+M_\et^2-2M_K^2)
\eeq
which is supposed to hold%
\footnote{For an explicit check on the lattice see Ref.\cite{Cichy:2015jra} and references therein.}
at leading order in the $1/\Nc$ expansion.
Interestingly, the right-hand side refers to full QCD quantities%
\footnote{We use the Bern normalization $F_\pi=f_\pi/\sqrt{2}$ of the pion decay constant, where $F_\pi^\mr{phys}=92.4(3)\MeV$ in QCD
with physical quark masses, and $F=86.2(5)\MeV$ in the 2-flavor chiral limit \cite{FlavourLatticeAveragingGroupFLAG:2021npn}.}
only, so the relation links two distinct theories.

In the continuum the topological susceptibility in a finite Euclidean volume $V$ may be defined as
\beq
\chi_\mr{top}=\int\<q(x)q(0)\>\;\rd^4x = \lim_{p^2\to0} \frac{1}{V}\int\<q(x)q(y)\>\,e^{\ri p(x-y)}\;\rd^4x\,\rd^4y
\label{def_toposusc_withcontact}
\eeq
where $q(x)$ is the topological charge density.
In this approach two limits are involved, zero virtuality ($p^2\to0$) and infinite volume ($V\to\infty$) in toroidal geometry.
Alternatively, one may use the definition
\beq
\chi_\mr{top}=\frac{\<q^2\>}{V} \qquad\mbox{with}\qquad q=\int q(x)\;\rd^4x
\label{def_toposusc_fromcharge}
\eeq
the global topological charge $q\in\mathbb{Z}$.
Again a limit $V\to\infty$ is required, but now one is restricted to $p^2=0$.
The two approaches are equivalent (up to a possible contact term \cite{Seiler:1987ig}).

On the lattice one may start from definitions analogous to (\ref{def_toposusc_withcontact}) or (\ref{def_toposusc_fromcharge}), but the renormalization details are different.
On a smoothed copy ($t>0$ in the terminology of Sec.~\ref{sec:smoothing}) of the gauge field
\bea
q_\mr{nai}(x)&=&\frac{1}{32\pi^2}\ep_{\mu\nu\rh\si}\mr{Tr}[F_{\mu\nu}(x)F_{\rh\si}(x)]
\nonumber
\\
&=&\frac{1}{4\pi^2}\mr{Tr}[F_{12}(x)F_{34}(x)-F_{13}(x)F_{24}(x)+F_{14}(x)F_{23}(x)]
\label{def_qnai}
\eea
is a ($t$-dependent) gluonic definition of the field strength tensor $F_{\mu\nu}(x)=F_{\mu\nu}(x)^aT^a$ with $T^a=\frac{1}{2}\la^a$ and thus of the \emph{local} topological charge density $q(x)$.
In this case the topological susceptibility
\beq
\chi_\mr{top}=Z_q^2(\be,t)\chi_\mr{nai}+M(\be,t) \qquad\mbox{with}\qquad \chi_\mr{nai}=\frac{a^4}{N}\sum_{x,y\in\Lambda}q_\mr{nai}(x)q_\mr{nai}(y)
\label{additive_renormalization}
\eeq
akin to (\ref{def_toposusc_withcontact}) renormalizes both multiplicatively and additively \cite{Campostrini:1989dh,DiGiacomo:1991ba,Alles:1997nu}.
Here $N=(L/a)^4$ is the number of lattice sites, while $V=L^4$ is the box volume in physical units.
Alternatively
\beq
q_\mr{ren}=\mr{round}(Z_q(\be,t)q_\mr{nai}) \qquad\mbox{with}\qquad q_\mr{nai}=a^4\sum_{x\in\Lambda}q_\mr{nai}(x)
\label{def_qren}
\eeq
is a \emph{global} topological charge wherein $q_\mr{nai}$ at $t>0$ is only multiplicatively%
\footnote{An additive renormalization of $q_\mr{nai}$ is excluded by the CP symmetry of the lattice theory.}
renormalized.
Based on $q_\mr{ren}$ defined in (\ref{def_qren}) on smoothed gauge fields, one may proceed to define the topological susceptibility
\beq
\chi_\mr{top}=\frac{\<q_\mr{ren}^2\>}{V}
\label{def_chitop}
\eeq
akin to (\ref{def_toposusc_fromcharge}), and no further renormalization is needed in this last step \cite{Hoek:1986nd,Lucini:2001ej,DelDebbio:2002xa,Durr:2006ky,Borsanyi:2015cka,Bonati:2014tqa,Alexandrou:2017hqw,Athenodorou:2020ani,Athenodorou:2021qvs}.
Put differently, the relative weights ${\cal Z}_q(V)/{\cal Z}(V)$ of the topological sectors in the $\theta=0$ YM partition function
\beq
{\cal Z}(V)=\sum_{q=-\infty}^\infty {\cal Z}_q(V)=\sum_{q=-\infty}^\infty \int e^{-S_\mr{glue}[U]} [\mathrm{D}U]_{q,V}
\label{def_partitionfunction}
\eeq
at $t>0$ qualify as observables, and $\<q^2\>$ is defined as the second moment of the partition function.
We will also address the higher moment combination $\<q^4\>-3\<q^2\>^2$, known as ``excess kurtosis''.

The remainder of this article is organized as follows.
In Sec.~\ref{sec:smoothing} we specify how we smooth the gauge configurations to define a variety of $q_\mr{nai}\in\mathbb{R}$ and $q_\mr{ren}\in\mathbb{Z}$ which is less susceptible to UV noise.
In Sec.~\ref{sec:setup} we discuss how we choose the bare parameters to generate a number of ensembles with a joint physical volume and decreasing lattice spacings,
and we give details of how we compute the $Z_q$-factors for the global topological charge (\ref{def_qren}).
In Sec.~\ref{sec:susc} our analysis is presented which yields the continuum limit of the topological susceptibility in a fixed physical volume,
together with a robust estimate of the theoretical uncertainty involved.
In Sec.~\ref{sec:kurt} the same type of analysis is repeated for the excess kurtosis of the global charge distribution.
In Sec.~\ref{sec:infvol} we use another set of simulations to study the infinite-volume behavior of the quantities studied in the previous two sections,
again with a careful estimate of the theoretical uncertainty involved.
In Sec.~\ref{sec:discussion} our result is compared to the literature, and Sec.~\ref{sec:conc} gives a summary and discusses some prospects for future research.

%%%%%%%%%%%%%%%%%%%%%%%%%%%%%%%%%%%%%%%%%%%%%%%%%%%%%%%%%%%%%%%%%%%%%%%%%%%%%%%

\section{Smoothing via stout smearing or Wilson flow \label{sec:smoothing}}

%%%%%%%%%%%%%%%%%%%%%%%%%%%%%%%%%%%%%%%%%%%%%%%%%%%%%%%%%%%%%%%%%%%%%%%%%%%%%%%

In contemporary lattice field theory two closely related smoothing schemes are used, stout smearing \cite{Morningstar:2003gk} and gradient flow \cite{Narayanan:2006rf,Luscher:2010iy,Luscher:2011bx}.
We use the output $V_\mu(x)$ of either one to define a clover operator
\bea
C_{\mu\nu}(x) &=& V_\mu(x) V_\nu(x+\hat{\mu}) V_\mu\dag(x+\hat{\nu}) V_\nu\dag(x) \nonumber\\
              &+& V_\nu(x) V_\mu\dag(x-\hat{\mu}+\hat{\nu}) V_\nu\dag(x-\hat{\mu}) V_\mu(x-\hat{\mu}) \nonumber\\
              &+& V_\mu\dag(x-\hat{\mu}) V_\nu\dag(x-\hat{\mu}-\hat{\nu}) V_\mu(x-\hat{\mu}-\hat{\nu}) V_\nu(x-\hat{\nu}) \nonumber\\
              &+& V_\nu\dag(x-\hat{\nu}) V_\mu(x-\hat{\nu}) V_\nu(x+\hat{\mu}-\hat{\nu}) V_\mu\dag(x)
\label{def_cmunu}
\eea
for a given lattice site $x\in\Lambda$.
Here $\hat{\mu}$ denotes $a$ times the unit vector in the direction $\mu$.
This $C_{\mu\nu}(x)$ is identified with $4I_{N_c}$ plus $4\ri$ times the field strength operator.
Unlike $F_{\mu\nu}(x)$ in the continuum%
\footnote{We like observables like the field strength $F_{\mu\nu}(x)$ or the gauge potential $A_\mu(x)$ in $U_\mu(x)=P\{\exp[\ri g\int_x^{x+\hat\mu}A(s)\,\rd s]\}$
to be hermitian quantities, as is common practice in quantum mechanics.}
the latter is not exactly hermitian (in color space) and exactly traceless.
Therefore we define
\beq
F_{\mu\nu}(x)=P_\mr{TH}[\frac{1}{4\ri}C_{\mu\nu}(x)] \qquad\mbox{with}\qquad P_\mr{TH}[M]=\frac{1}{2}(M+M\dag)-\frac{1}{2N_c}\mr{Tr}(M+M\dag)\,I_{N_c}
\eeq
as the traceless hermitian part of (\ref{def_cmunu}) divided by $4\ri$.

Our smeared field $V_\mu(x)$ emerges from the unsmeared $U_\mu(x)$ through $n$ steps of stout%
\footnote{Alternatively, one may remove the $\ri$ and $1/\ri$ in (\ref{def_stout}, \ref{def_techn}) and replace $P_\mr{TH}$ by the traceless antihermitian projector.}
smearing
\beq
V_\mu(x)=V^{(n)}_\mu(x) \;,\qquad V^{(n)}_\mu(x)=e^{\ri \rho Q_\mu^{(n-1)}(x)}\,V_\mu^{(n-1)}(x) \;,\qquad V_\mu^{(0)}(x)=U_\mu(x)
\label{def_stout}
\eeq
where the stout parameter should be chosen in the interval $0<\rh<0.125$ in 4D \cite{Capitani:2006ni}.
The operator
\beq
Q_\mu^{(n-1)}(x) = P_\mr{TH}\Big[\frac{1}{\ri} S_\mu^{(n-1)}(x) V^{(n-1)\,\dagger}(x)\Big]
\label{def_techn}
\eeq
contains the product of $V_\mu(x)$ and $S_\mu\dag(x)$ which also appears in the Wilson gauge action, with
\beq
S_\mu^{(k)}(x) = \sum_{\nu\neq\mu} \Big\{
V_\nu^{(k)}(x) V_\mu^{(k)}(x+\hat{\nu}) V_\nu^{(k)\dagger}(x+\hat{\mu}) +
V_\nu^{(k)\dagger}(x-\hat{\nu})V_\mu^{(k)}(x-\hat{\nu})V_\nu^{(k)}(x+\hat{\mu}-\hat{\nu})
\Big\}
\eeq
being the staple around the link $V_\mu^{(k)}(x)$, pointing in the same direction as the link itself.

The main advantage of stout smearing is that one stays in the gauge group, hence no ``backprojection'' to $SU(3)$ is needed.
This is the technical basis of the suggestion made in Refs.~\cite{Narayanan:2006rf,Luscher:2010iy,Luscher:2011bx} to consider the limit $\rho\to0$ and $n\to\infty$ where the product $n\times\rho=t/a^2$ is kept constant.
The quantity $t/a^2$ is called the ``flow time in lattice units'' and has the meaning of a cumulative sum of the $\rh$-parameters used in all steps.
Hence, to reach $t/a^2=0.84$ one may factor the sum as $7\times0.12$ or $14\times0.06$ or $28\times0.03$, and so on.
In this sequence the step-size error in the flow time decreases (ideal Wilson flow means zero step size).
For some applications (e.g.\ measuring $t_0$ or $w_0$ \cite{Luscher:2010iy,Luscher:2011bx,BMW:2012hcm}) it is important to keep the flow time discretization effect small.
For other applications (e.g.\ the one we have in mind) it is less important (we shall come back to this point in Sec.~\ref{sec:setup}).
The conceptual link between stout smearing and (ideal) Wilson flow has been discussed in Refs.~\cite{Luscher:2011bx,Nagatsuka:2023jos,Ammer:2024hqr}.

In today's lattice literature the main difference between ``stout smearing'' and ``gradient flow'' is the quantity which is held fixed in the continuum limit $a\to0$.
The terminology ``stout smearing'' usually implies that $\rho$ and $n$, and hence the flow time $t/a^2$ in \emph{lattice units} is kept constant at all $\be$.
With this strategy the naive topological charge density (\ref{def_qnai}) is an ultralocal operator (fixed footprint in lattice units) at all $\be$, and becomes pointlike in the limit $a\to0$.
The terminology ``gradient flow'' is usually chosen when the flow time $t/r_0^2=t/a^2\times(a/r_0)^2$ in \emph{physical units} is held fixed at all $\be$.
In our approximation (via stout smearings with $\rho$ fixed) the number of steps is then bound to increase like $(r_0/a)^2$ toward the continuum.
As a result, $q_\mr{nai}(x)$ is ``regularized'' over a distance $\sqrt{8t}$ (which is a fixed distance in $r_0$ units) and no longer ultralocal \cite{Luscher:2010iy}.
Hence, in this strategy a second regulator is in place, which persists in the continuum limit $a\to0$, see e.g.\ Ref.~\cite{Ammer:2024hqr} for a discussion.

In Refs.~\cite{Ce:2015qha,Bonanno:2023ple} it is stated that for the topological susceptibility $\chi_\mr{top}$ the continuum limit
can be taken at a fixed flow time in physical units, too, not just at a fixed flow time in lattice units (as was traditionally done).
In other words, for this observable the additional flow regulator does not introduce any systematic bias
(the continuum extrapolated value $\chi_\mr{top}r_0^4$ at fixed $t/r_0^2$ would be the same as, say, with $7$ stout steps),
provided $\sqrt{t}$ is in a range where it smoothes out short-range fluctuations but leaves long-range fluctuations (say for $r>r_0$) unaffected.
In this article we aim for a numerical comparison of the two strategies.

%%%%%%%%%%%%%%%%%%%%%%%%%%%%%%%%%%%%%%%%%%%%%%%%%%%%%%%%%%%%%%%%%%%%%%%%%%%%%%%

\section{Lattice setup and renormalization factors \label{sec:setup}}

%%%%%%%%%%%%%%%%%%%%%%%%%%%%%%%%%%%%%%%%%%%%%%%%%%%%%%%%%%%%%%%%%%%%%%%%%%%%%%%

Our goal is to set up a series of lattice simulations for $\chi_\mr{top}$ with a fixed physical volume $V$.
We choose $V=(2.4783\,r_0)^4$ which is about 51\% larger than the volume $V=(2.2356\,r_0)^4$ of Ref.~\cite{Durr:2006ky}.
We use the Wilson gauge action, and Eq.~(14) of Ref.~\cite{Durr:2006ky} parametrizes $r_0/a$ as a function of $\be$.
Today, $t_0$ \cite{Luscher:2010iy,Luscher:2011bx} and $w_0$ \cite{BMW:2012hcm} are valuable alternatives to set the scale, but previous results refer to $r_0$,
and we see no reason to believe that the continuum limit would be different%
\footnote{The quantity $t_0$ has the dimension of an area (or inverse mass squared), hence the ratio $\sqrt{t_0}/r_0$ is a real number
which assumes a universal value in the continuum limit $a\to0$.}
at all.

\begin{table}[tb]
\centering
\begin{tabular}{|cccc|ccc|}
\toprule
$L/a$ & $\beta$ & $r_0/a$ & $a\,[\mr{fm}]$ & 7\,stout & flow\,0.21\,fm & flow\,0.30\,fm \\
\midrule
% 4.8423   0.0982(13)  fm
% 5.6493   0.0842(11)  fm
% 6.4561   0.07368(99) fm
% 7.2632   0.06550(88) fm
% 8.0695   0.05895(79) fm
% 9.6836   0.04912(66) fm
% 11.298   0.04211(57) fm
12 & 5.9421 & 4.842 & 0.098 & $7 \times 0.12$ & $ 9 \times 0.06   =0.54 $ & $ 9 \times 0.12  =1.08$ \\
14 & 6.0314 & 5.649 & 0.084 & $7 \times 0.12$ & $12 \times 0.06125=0.735$ & $12 \times 0.1225=1.47$ \\
16 & 6.1142 & 6.456 & 0.074 & $7 \times 0.12$ & $16 \times 0.06   =0.96 $ & $16 \times 0.12  =1.92$ \\
18 & 6.1912 & 7.263 & 0.066 & $7 \times 0.12$ & $20 \times 0.06075=1.215$ & $20 \times 0.1215=2.43$ \\
20 & 6.2629 & 8.070 & 0.059 & $7 \times 0.12$ & $25 \times 0.06   =1.5  $ & $25 \times 0.12  =3.00$ \\
24 & 6.3929 & 9.684 & 0.049 & $7 \times 0.12$ & $36 \times 0.06   =2.16 $ & $36 \times 0.12  =4.32$ \\
28 & 6.5079 & 11.30 & 0.042 & $7 \times 0.12$ & $49 \times 0.06   =2.94 $ & $49 \times 0.12  =5.88$ \\
\bottomrule
\end{tabular}
\caption{\sl Overview of the box sizes and couplings selected for the continuum scaling analysis.
The volume $V=(2.4783\,r_0)^4$ is fixed in physical units, based on $r_0/a$ as given in Eq.~(14) of Ref.~\cite{Durr:2006ky}.
The ``7\,stout'' smoothing strategy keeps the flow time in lattice units fixed at $t/a^2=7\times0.12=0.84$,
tantamount to $\sqrt{8t}\simeq2.59\,a\to0$ in physical units for $\be\to\infty$.
The ``flow\,0.21\,fm'' strategy sets the flow time to $t/a^2=(N/4)^2\times0.06$, tantamount to $\sqrt{8t}=0.429\,r_0\simeq0.21\,\fm$.
The ``flow\,0.30\,fm'' strategy sets the flow time to $t/a^2=(N/4)^2\times0.12$, tantamount to $\sqrt{8t}=0.607\,r_0\simeq0.30\,\fm$.
\label{tab:smear_data}}
\end{table}

Details of our plan are presented in Tab.~\ref{tab:smear_data}.
We use three smoothing strategies, named ``7\,stout'' (fixed $t/a^2=0.84$) and ``0.21\,fm'', ``0.30\,fm'' (fixed $t/r_0^2$), respectively.
To get statistically independent continuum extrapolated values, we shall generate $7\times3=21$ independent%
\footnote{The original plan was to generate only five lattice spacings. At $\be=6.0314,6.1912$ a slight
increase of $\rho$ from 0.06 to $\sim0.061$ or from 0.12 to $\sim0.122$ was needed to fit in these lattices.}
ensembles.
The flow time discretization effects in the latter two strategies are expected%
\footnote{Comparing the top-left and bottom-left panels in Fig.~13 of Ref.~\cite{Alexandrou:2017hqw} we see no hint
for any flow time discretization effects in $\chi_\mr{top}^{1/4}r_0$ at flow times and step sizes similar to ours.}
to be small.

\begin{table}[tb]
\centering
\begin{tabular}{|cc|ccc|ccc|}
\toprule
$L/a$ & $\beta$ & 7\,stout & flow\,0.21\,fm & flow\,0.30\,fm & 7\,stout & flow\,0.21\,fm & flow\,0.30\,fm \\
\midrule
12 & 5.9421 & 100000[10] &  100000[10] &  100000[10] & 0.650(10) & 0.640(10) & 0.660(10) \\
14 & 6.0314 & 209519[10] &  200000[10] &  200000[10] & 1.120(20) & 1.130(20) & 1.150(20) \\
16 & 6.1142 &  40851[64] &   50654[64] &   49617[64] & 0.570(10) & 0.560(10) & 0.550(10) \\
18 & 6.1912 & 310191[81] &   54760[81] &   54577[81] & 0.730(20) & 0.710(20) & 0.730(20) \\
20 & 6.2629 & 78749[100] &  83474[100] & 106925[100] & 1.020(2)  & 1.060(30) & 1.030(20) \\
24 & 6.3929 & 93292[100] &  56238[144] &  69276[144] & 3.08(10)  & 2.209(90) & 2.280(80) \\
28 & 6.5079 & 68641[196] & 126804[196] &  87726[196] & 4.81(24)  & 5.04(26)  & 5.34(28)  \\
\bottomrule
\end{tabular}
\caption{\sl Details of the ensembles used in the continuum extrapolation.
Columns three to five contain the number of measurements and the number of update packages between adjacent measurements in the format $n_\mr{meas}[n_\mr{sepa}]$,
a separate stream was generated for each smoothing strategy.
The last three columns contain $\tau_\mr{int}(q_\mr{ren}^2)$ where $q_\mr{ren}$ uses the smoothing strategy listed in the column head.
\label{tab:sim_data}}
\end{table}

Tab.~\ref{tab:sim_data} lists the ensembles generated, along with the number of measurements made and the number of update packages between adjacent measurements.
An update package consists of a heat bath sweep \cite{Creutz:1980zw,Cabibbo:1982zn,Fabricius:1984wp,Kennedy:1985nu} followed by four overrelaxation sweeps \cite{Adler:1981sn,Creutz:1987xi,Brown:1987rra}.
Occasionally, a $P$-transformation \cite{Leinweber:2003sj} of the gauge configuration is applied (this preserves the action and flips the sign of $q_\mr{ren}$).
The last three columns give $\tau_\mr{int}(q_\mr{ren}^2)$, where $q_\mr{ren}$ uses the smoothing strategy listed in the column head.

\begin{figure}[tb]
\includegraphics[width=0.52\textwidth]{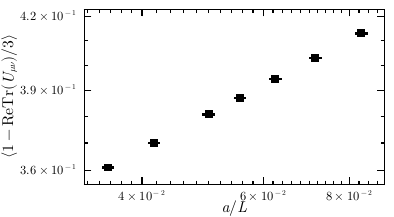}%
\includegraphics[width=0.48\textwidth]{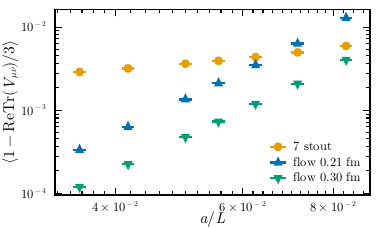}%
\vspace*{-4pt}
\caption{\sl $\<1-\mr{Re}\,\mr{Tr}(U_{\mu\nu})/3\>$ of the ensembles used in the continuum extrapolation, unsmeared (left)
and with one of the three smoothing strategies (right).
\label{fig:oneminusplaq}}
\end{figure}

The original (unsmeared) plaquettes are displayed in the first panel of Fig.~\ref{fig:oneminusplaq}.
In log-log representation versus $a/r_0$ (or our $a/L$) they appear almost linear.
A marked difference between the ``7\,stout'' smoothing strategy on the one hand and the ``flow\,0.21\,fm'', ``flow\,0.30\,fm'' strategies on the other hand is illustrated in the second panel.
With a fixed flow time in lattice units (``7\,stout'') the slope in the log-log representation is small, while with a fixed flow time in physical units (``flow\,0.21\,fm'', ``flow\,0.30\,fm'') it is much steeper.
This is unsurprising, since with the latter two strategies the number of stout steps in Tab.~\ref{tab:smear_data} proliferates $\propto(r_0/a)^2$ toward the continuum.

\begin{table}[tb]
\centering
\begin{tabular}{|cc|ccc|}
\toprule
$L/a$ & $\beta$ & 7\,stout & flow\,0.21\,fm & flow\,0.30\,fm \\
\midrule
12 & 5.9421 & 1.2757(18)  & 1.3782(57)  & 1.2337(20)  \\
14 & 6.0314 & 1.22881(57) & 1.2522(12)  & 1.15821(41) \\
16 & 6.1142 & 1.1974(10)  & 1.17807(80) & 1.11444(49) \\
18 & 6.1912 & 1.17499(62) & 1.13280(43) & 1.08579(26) \\
20 & 6.2629 & 1.15682(53) & 1.10237(47) & 1.06672(17) \\
24 & 6.3929 & 1.13312(31) & 1.06700(16) & 1.04426(11) \\
28 & 6.5079 & 1.11818(37) & 1.04740(10) & 1.03149(12) \\
\bottomrule
\end{tabular}
\caption{\sl The multiplicative renormalization factor $Z_q$ of the gluonic topological charge,
determined for each smoothing strategy at various lattice spacings in a fixed physical volume $V = (2.4783\,r_0)^4$.
\label{tab:Z_factor}}
\end{table}

% Mir ist aufgefallen, dass ich dir die P-Werte von dem linken Z_q plot noch nicht gegeben habe. Die sind:
% "7 Stout": P=0.341 mit Limit 1.07531(40) fuer a\to0
% "0.21 fm": P=0.732 mit Limit 1.0000(15)  fuer a\to0
% "0.30 fm": P=0.160 mit Limit 0.99957(55) fuer a\to0

\begin{figure}[tb]
\centering
\includegraphics[width=1.0\textwidth]{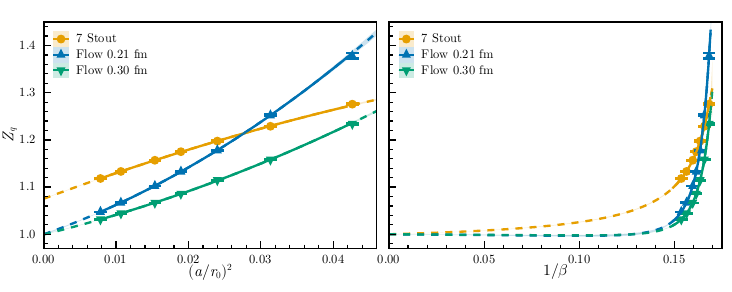}%
\vspace*{-12pt}
\caption{\label{fig:Z_scaling_pub}\sl
The $Z_q$ factors involved, with quadratic fits in $(a/r_0)^2$ (left) and rational fits in $g_0^2$ (right).}
\vspace*{+12pt}
\end{figure}

Finally, we need a nonperturbative determination of the renormalization factors $Z_q(\be,t)$.
Following Refs.~\cite{DelDebbio:2002xa,Durr:2004xu,Durr:2006ky,Bonati:2015sqt} we calculate, for each ensemble and smoothing strategy, the quantity
\beq
\chi_\mr{min}^2=\min_{1\leq Z \leq2} \sum_{i=1}^{n_\mr{conf}}\Big(Z q_\mr{nai}^{(i)}-\mr{round}(Z q_\mr{nai}^{(i)})\Big)^2
\eeq
and the $Z$ which realizes the minimum is (for the given $\be$ and smoothing recipe) the global topological charge renormalization factor $Z_q$ in Eq.~(\ref{def_qren}).
The results are tabulated in Tab.~\ref{tab:Z_factor} and displayed in Fig.~\ref{fig:Z_scaling_pub}.
For each smoothing strategy the $Z_q$ factor decreases toward the continuum.
Plotting $Z_q$ as a function of $a^2$ (left panel) the function seems to pass through $1$ at $a=0$ for the physical flow time strategies
(``flow\,0.21\,fm'', ``flow\,0.30\,fm''), even though these were unconstrained fits.
If the flow time is fixed in lattice units (``7\,stout''), the extrapolation in $(a/r_0)^2$ misses $1$ in the continuum by many sigmas.
The theoretically better motivated extrapolation in $g_0^2=6/\beta$ (right panel) is more cumbersome to fit.
Eventually, we were successful with the ansatz $Z_q=(1+a_1/\beta+a_2/\beta^2)/(1+b_1/\beta+b_2/\beta^2)$.
With $\mr{dof}=7-4=3$ it yields the $P$ values $0.171$, $0.896$ and $0.350$ for the ``7\,stout'', ``flow\,0.21\,fm'' and ``flow\,0.30\,fm'' strategies, respectively.

%%%%%%%%%%%%%%%%%%%%%%%%%%%%%%%%%%%%%%%%%%%%%%%%%%%%%%%%%%%%%%%%%%%%%%%%%%%%%%%

\section{Continuum analysis for the topological susceptibility \label{sec:susc}}

%%%%%%%%%%%%%%%%%%%%%%%%%%%%%%%%%%%%%%%%%%%%%%%%%%%%%%%%%%%%%%%%%%%%%%%%%%%%%%%

With these preparatory steps completed, we are in a position to present the analysis for the continuum extrapolation of the topological susceptibility.

\begin{table}[tb]
\centering
\begin{tabular}{|cc|ccc|ccc|}
\toprule
$L/a$ & $\beta$ & 7\,stout & flow\,0.21\,fm & flow\,0.30\,fm & 7\,stout & flow\,0.21\,fm & flow\,0.30\,fm \\
\midrule
12 & 5.9421 & 1.4653(78) & 1.2158(64) & 1.5970(85) & 2.453(12)  & 2.387(12) & 2.486(13) \\
14 & 6.0314 & 1.5362(74) & 1.4734(73) & 1.7419(88) & 2.369(11)  & 2.364(12) & 2.365(12) \\
16 & 6.1142 & 1.554(12)  & 1.633(11)  & 1.806(11)  & 2.268(17)  & 2.294(14) & 2.254(14) \\
18 & 6.1912 & 1.5888(58) & 1.715(13)  & 1.871(14)  & 2.2154(81) & 2.214(16) & 2.212(17) \\
20 & 6.2629 & 1.619(12)  & 1.781(13)  & 1.897(12)  & 2.185(16)  & 2.171(16) & 2.162(14) \\
24 & 6.3929 & 1.617(19)  & 1.834(24)  & 1.899(22)  & 2.088(25)  & 2.090(27) & 2.072(24) \\
28 & 6.5079 & 1.627(28)  & 1.847(24)  & 1.926(32)  & 2.046(35)  & 2.027(26) & 2.050(34) \\
\bottomrule
\end{tabular}
\caption{\sl
Ensemble average and statistical error of $\<q_\mr{nai}^2\>$ (columns three to five) and $\<q_\mr{ren}^2\>$ (last three columns),
as determined for each smoothing strategy and lattice spacing in the fixed physical volume $V = (2.4783\,r_0)^4$.
These results reflect $7\times3$ different ensembles.
\label{tab:susc_data}}
\end{table}

Our $7\times3=21$ ensembles with a joint physical volume are used to measure the naive topological charge (\ref{def_qnai}) and, based on the $Z_q$ listed in Tab.~\ref{tab:Z_factor}, the renormalized charge (\ref{def_qren}).
In either case the second moment of the distribution is determined, and the resulting $\<q_\mr{nai}^2\>$ (first three columns) and $\<q_\mr{ren}^2\>$ (last three columns) are listed in Tab.~\ref{tab:susc_data}.
Given (\ref{def_chitop}) and $V=(2.4783\,r_0)^4$, these numbers must be divided by $2.4783^4$ to obtain $\chi_\mr{top}r_0^4$.
All that remains to be done is to get rid of the discretization effects by means of a continuum extrapolation with $O(a^2)$ cutoff effects.

\begin{figure}[tb]
\includegraphics[width=1.0\textwidth]{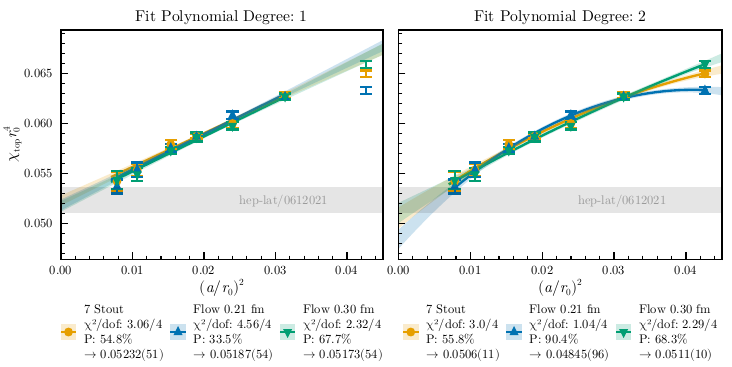}%
\vspace*{-4pt}
\caption{\sl Continuum extrapolation of the topological susceptibility with an ansatz linear in $a^2$ based on the six finest
spacings (left) and with an ansatz quadratic in $a^2$ based on all seven spacings (right).
\label{fig:chir04_compare_puball}}
\end{figure}

\begin{figure}[tb]
\includegraphics[width=1.0\textwidth]{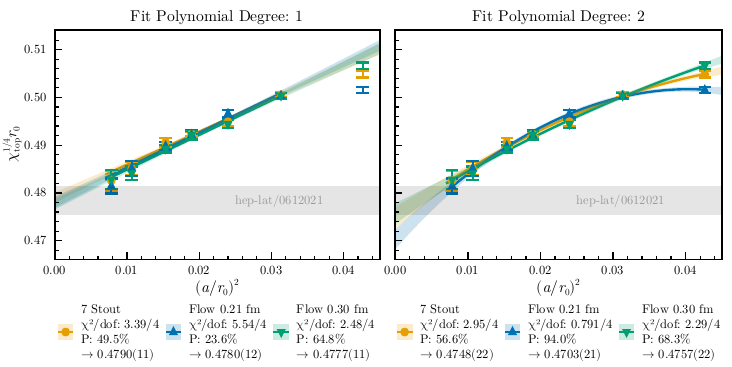}%
\vspace*{-4pt}
\caption{\sl Same as Fig.~\ref{fig:chir04_compare_puball} but with the quantity $\chi_\mr{top}^{1/4}r_0$ on the ordinate (and six freshly created fits).
\label{fig:chiqrt_compare_puball}}
\end{figure}

First we discuss this extrapolation for the data based on $q_\mr{ren}^2$ (last three columns of Tab.~\ref{tab:susc_data}),
since this is the standard procedure \cite{Hoek:1986nd,Lucini:2001ej,DelDebbio:2002xa,Durr:2006ky,Borsanyi:2015cka,Bonati:2014tqa,Alexandrou:2017hqw,Athenodorou:2020ani,Athenodorou:2021qvs}.
We obtain good fits with a ``const+linear'' fit (in $a^2$) that excludes the coarsest ($\be=5.9421$) lattice spacing, and with a ``const+linear+quadratic'' fit that includes all lattice spacings.
These fits are shown in Fig.~\ref{fig:chir04_compare_puball}.
The extrapolated values at $a=0$ for the three smoothing strategies are statistically independent (as are the ensembles).
With the linear ansatz they agree very closely, with the quadratic ansatz (in $a^2$) there is a visible spread, but they still agree within statistical errors.
Extrapolating $\chi_\mr{top}r_0^4$ is not the only possibility, also $\chi_\mr{top}^{1/4}r_0$ or other powers are permissible.
We add the latter option to control the pertinent systematics; the results are shown in Fig.~\ref{fig:chiqrt_compare_puball}.
Again ``const+linear'' without the coarsest lattice spacing and ``const+linear+quadratic'' with all data included are found to yield acceptable fits.

\begin{table}[tb]
\centering
\begin{tabular}{|c|cc|cc|}
\toprule
 & $[\chi_\mr{top}r_0^4]_\mr{lin.}$ & $[\chi_\mr{top}r_0^4]_\mr{quad.}$ & $[\chi_\mr{top}^{1/4}r_0]_\mr{lin.}$ & $[\chi_\mr{top}^{1/4}r_0]_\mr{quad.}$ \\
\midrule
7\,stout       & 0.05232(51) & 0.0506(11)  & 0.4790(11) & 0.4748(22) \\
flow\,0.21\,fm & 0.05187(54) & 0.04845(96) & 0.4780(12) & 0.4703(21) \\
flow\,0.30\,fm & 0.05173(54) & 0.0511(10)  & 0.4777(11) & 0.4757(22) \\
\bottomrule
\end{tabular}
\caption{\sl
Continuum value of $\chi_\mr{top}r_0^4$ and its fourth root, in a fixed physical volume $V = (2.4783\,r_0)^4$,
extracted with one of two fitting ans\"atze and one of three smearing strategies.
The statistical errors are highly correlated along each line, but fully uncorrelated along each column.
\label{tab:continuum_values}}
\end{table}

We have, for each one of the three smoothing strategies, two continuum extrapolations of the topological susceptibility and two of its fourth root.
These results are shown in Tab.~\ref{tab:continuum_values}.
For a given smoothing strategy the four entries describe a joint continuum limit (their spread indicates a systematic uncertainty).
It is, a priori, not clear that the ``ultralocal'' strategy (``7\,stout'') and either one of the ``fixed physical flow time'' strategies
(``flow\,0.21\,fm'' and ``flow\,0.30\,fm'') would yield the same continuum limit.
As mentioned in Sec.~\ref{sec:setup}, the latter two strategies introduce a second regulator which persists in the continuum limit.
In Refs.~\cite{Ce:2015qha,Bonanno:2023ple} it is stated that with a reasonable choice of $\sqrt{8t}$ the second regulator leaves a negligible imprint on $\chi_\mr{top}^{1/4}r_0$ at $a\to0$.
We are thus left with the task to condense, for each smoothing strategy, the four entries in Tab.~\ref{tab:continuum_values} into a single number
(with statistical and systematic uncertainties) to see whether our data support this statement.

%%% erste Zeile wird zu: 0.05202(61)(152) = [0.4776(14)(35)]^4    0.4782(13)(40)    0.4779(14)(37)(02) laut meiner Berechnung

\begin{table}[tb]
\centering
%%%cen=[ 0.05230 , 0.0506 ;0.05187,0.04845;0.05173,0.0511; 0.4789 , 0.4749 ;0.4780,0.4703;0.4777,0.4757]
%%%err=[ 0.00057 , 0.0011 ;0.00054,0.00096;0.00054,0.0010; 0.0012 , 0.0022 ;0.0012,0.0021;0.0011,0.0022]
%  cen=[ 0.05232 , 0.0506 ;0.05187,0.04845;0.05173,0.0511; 0.4790 , 0.4748 ;0.4780,0.4703;0.4777,0.4757]
%  err=[ 0.00051 , 0.0011 ;0.00054,0.00096;0.00054,0.0010; 0.0011 , 0.0022 ;0.0012,0.0021;0.0011,0.0022]
%w=(1./err.^2)./repmat(sum(1./err.^2,2),1,2)
%avg=sum(w.*cen,2)
%sta=sum(w.*err,2) %%% note: not sqrt(sum(err.^2,2))
%sys=abs(cen(:,1)-cen(:,2)).*erf(abs(cen(:,1)-cen(:,2))./sqrt(2)./max(err,[],2))
%com=sqrt(sta.^2+sys.^2)
\begin{tabular}{|c|ccc|}
\toprule
 & $\chi_\mr{top}r_0^4$ & $\chi_\mr{top}^{1/4}r_0$ & combined \\
%(0.4782-0.4776)*erf((0.4782-0.4776)/sqrt(2)/max(0.0035,0.0040))=0.0001
%(0.4761-0.4753)*erf((0.4761-0.4753)/sqrt(2)/max(0.0080,0.0077))=0.0001
%(0.4773-0.4766)*erf((0.4773-0.4766)/sqrt(2)/max(0.0007,0.0013))=0.0003
\midrule
7\,stout       & 0.05202(61)(152)=$[0.4776(14)(35)]^4$ & 0.4782(13)(40) & 0.4779(14)(37)(01) \\
flow\,0.21\,fm & 0.05105(64)(342)=$[0.4753(15)(80)]^4$ & 0.4761(14)(77) & 0.4759(15)(79)(01) \\
flow\,0.30\,fm & 0.05159(64)(030)=$[0.4766(15)(07)]^4$ & 0.4773(13)(13) & 0.4769(14)(10)(03) \\
\bottomrule
\end{tabular}
\caption{\sl
Results of the continuum extrapolation of $\chi_\mr{top}r_0^4$ (left column) or $\chi_\mr{top}^{1/4}r_0$ (middle column).
\label{tab:results}}
\end{table}

We proceed in two steps (first we combine the two fit ans\"atze, then the two ordering options of the root and the extrapolation).
For the first line of Tab.~\ref{tab:continuum_values} this means that we combine $0.05230(57)_\mr{stat}$ and $0.0506(11)_\mr{stat}$ to become $0.05194(68)_\mr{stat}$,
where the average of the statistical errors was done with the same weights that were used for the average of the central values, since these errors are highly correlated.
In addition, the combined value needs to be attributed a systematic uncertainty to reflect the spread between the two numbers from which it was derived.
For combining two central values $c^{(1)},c^{(2)}$ with statistical uncertainties $\si_\mr{stat}^{(1)},\si_\mr{stat}^{(2)}$
Refs.~\cite{ExtendedTwistedMassCollaborationETMC:2022sta,Bonanno:2023xkg} recommend using
\beq
\sigma_\mr{syst} = |c^{(1)}-c^{(2)}| \; \mr{erf}\Big(\frac{|c^{(1)}-c^{(2)}|}{\sqrt{2}\max(\sigma_\mr{stat}^{(1)},\sigma_\mr{stat}^{(2)})}\Big)
\label{eq:recipe_systerror}
\eeq
since this is the difference between the extrapolations with the two fitting ans\"atze, multiplied by the probability that this difference is due to a statistical fluctuation.
For the ``7\,stout'' strategy this yields $\chi_\mr{top}r_0^4=0.05202(61)_\mr{stat}(152)_\mr{syst}$ in the continuum,
and $\chi_\mr{top}^{1/4}r_0=0.4782(13)_\mr{stat}(40)_\mr{syst}$.
Combining the latter two values by the same recipe yields $0.4779(14)_\mr{stat}(37)_\mr{syst}(01)_\mr{syst}$ in the continuum.
This number is stored in the last column of Tab.~\ref{tab:results}, along with the ``flow\,0.21\,fm'' and ``flow\,0.30\,fm'' counterparts.
These three figures are consistent within (overall) errors, in line with Refs.~\cite{Ce:2015qha,Bonanno:2023ple}.

\begin{figure}[tb]
\centering
\includegraphics[width=0.8\textwidth]{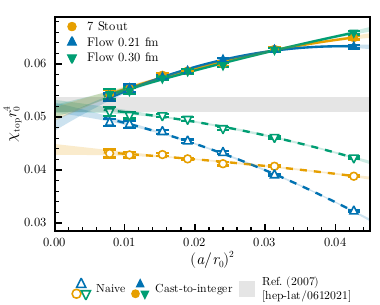}%
\vspace*{-4pt}
\caption{\sl The standard quadratic-fit extrapolation as shown in the right panel of Fig.~\ref{fig:chir04_compare_puball} (full lines connect $\<q_\mr{ren}^2\>/V$)
is compared to another approach where the $Z_q$-factor and the cast-to-integer operation are omitted (dashed lines connect $\<q_\mr{nai}^2\>/V$).
In both cases the data for all three smearing strategies are shown.
The most precise determination to date, Ref.~\cite{Durr:2006ky}, is included for comparison.
\label{fig:chir04_nrall}}
\end{figure}

This nice agreement among the smoothing strategies at the level of the continuum result contrasts with a technical difference that was alluded to in Fig.~\ref{fig:Z_scaling_pub} already.
Based on the extrapolation in the left panel, both $Z_q^\mr{0.21\,fm}$ and $Z_q^\mr{0.30\,fm}$ seem to tend to $1$ in the limit $a\to0$, while $Z_q^\mr{7\,stout}$ does not.
This suggests that one might obtain the correct continuum limit for $\chi_\mr{top}r_0^4$ based on $\<q_\mr{nai}^2\>$, that is \emph{without} the $Z_q$-factor inherent in $\<q_\mr{ren}^2\>$, but only for the fixed-flowtime strategies.
Such an extrapolation of $\chi_\mr{top}r_0^4$, based on the naive topological charge (\ref{def_qnai}), is shown with dashed lines in Fig.~\ref{fig:chir04_nrall}.
The standard extrapolation, based on the renormalized topological charge (\ref{def_qren}), is shown with full lines.
We observe that the fixed-flowtime strategies work with both $\<q_\mr{nai}^2\>$ and $\<q_\mr{ren}^2\>$.
By contrast, the ``7\,stout'' strategy works only with $\<q_\mr{ren}^2\>$, and the difference between the correct continuum limit (with $Z_q$) and the incorrect one (without), roughly $0.052-0.044\simeq0.008$,
reflects the additive renormalization constant $M$ in (\ref{additive_renormalization}) which is due whenever $\chi_\mr{top}$ is calculated from a naive charge at zero flowtime.
Near the continuum the ``flow\,0.21\,fm'' and ``flow\,0.30\,fm'' curves in Fig.~\ref{fig:chir04_nrall} are found to be flatter \emph{without} the $Z_q$-factors, and this leads to \emph{smaller} statistical errors in the continuum.
This suggests that our final results in Tab.~\ref{tab:results} come with conservatively assessed systematics.

To summarize, we note that the three smoothing strategies yield perfectly consistent results, in line with Refs.~\cite{Ce:2015qha,Bonanno:2023ple}.
In view of this, we select the ``0.30\,fm'' quote in Tab.~\ref{tab:results} as our final result (or average over the three strategies, it hardly makes any difference).
As we shall see in Sec.~\ref{sec:infvol}, this final result of the continuum extrapolation is subject to a tiny finite-volume shift.

%%%%%%%%%%%%%%%%%%%%%%%%%%%%%%%%%%%%%%%%%%%%%%%%%%%%%%%%%%%%%%%%%%%%%%%%%%%%%%%

\section{Continuum analysis for the topological excess kurtosis \label{sec:kurt}}

%%%%%%%%%%%%%%%%%%%%%%%%%%%%%%%%%%%%%%%%%%%%%%%%%%%%%%%%%%%%%%%%%%%%%%%%%%%%%%%

\begin{table}[tb]
\centering
\begin{tabular}{|c|@{\hskip 4pt}c@{\hskip 4pt}c@{\hskip 4pt}c@{\hskip 4pt}|}
\toprule
$L/a$ & $\<q_\mr{ren}^4\>/\<q_\mr{ren}^2\>^2-3$ & $\<q_\mr{ren}^4\>/\<q_\mr{ren}^2\>-3\<q_\mr{ren}^2\>$ & $\<q_\mr{ren}^4\>-3\<q_\mr{ren}^2\>^2$  \\
\midrule
12 & 0.135(20),0.088(18),0.129(18) & 0.331(48),0.209(43),0.320(45) & 0.81(12),  0.50(10),  0.80(11)   \\ % 5.9421 &
14 & 0.105(14),0.111(15),0.131(15) & 0.249(33),0.262(34),0.310(35) & 0.590(78), 0.620(82), 0.733(84)  \\ % 6.0314 &
16 & 0.098(29),0.138(21),0.116(20) & 0.221(65),0.317(49),0.261(44) & 0.50(15),  0.73(11),  0.587(100) \\ % 6.1142 &
18 & 0.120(12),0.093(24),0.130(25) & 0.265(26),0.205(52),0.287(56) & 0.586(57), 0.45(12),  0.64(12)   \\ % 6.1912 &
20 & 0.127(22),0.148(24),0.131(19) & 0.277(48),0.320(52),0.283(41) & 0.61(10),  0.69(11),  0.612(90)  \\ % 6.2629 &
24 & 0.131(34),0.136(35),0.075(32) & 0.273(71),0.283(73),0.155(67) & 0.57(15),  0.59(15),  0.32(14)   \\ % 6.3929 &
28 & 0.112(47),0.157(36),0.249(48) & 0.229(96),0.318(74),0.51(10)  & 0.47(20),  0.64(15),  1.03(21)   \\ % 6.5079 &
\bottomrule
\end{tabular}
\caption{\sl
Ensemble average and statistical error of the excess kurtosis varieties on all ensembles in a physical volume $V = (2.4783\,r_0)^4$.
Results for the ``7\,stout'', ``flow\,0.21\,fm'' and ``flow\,0.30\,fm'' smoothing strategies are comma separated in each cell.
The pertinent $\be$ are the same es in Tab.~\ref{tab:susc_data}.
\label{tab:kurt_data}}
\end{table}

% Die neuen Spalten mit <q_ren^4> fuer Tabelle 7 wuerden wie folgt lauten:
% 18.82(23), 17.59(21), 19.33(23)
% 17.43(18), 17.38(19), 17.51(19)
% 15.87(29), 16.47(25), 15.82(22)
% 15.40(27), 15.16(25), 15.32(26)
% 14.92(24), 14.81(25), 14.63(20)
% 13.65(35), 13.67(39), 13.18(34)
% 13.02(48), 12.92(37), 13.61(50)

\begin{figure}[tb]
\includegraphics[width=1.0\textwidth]{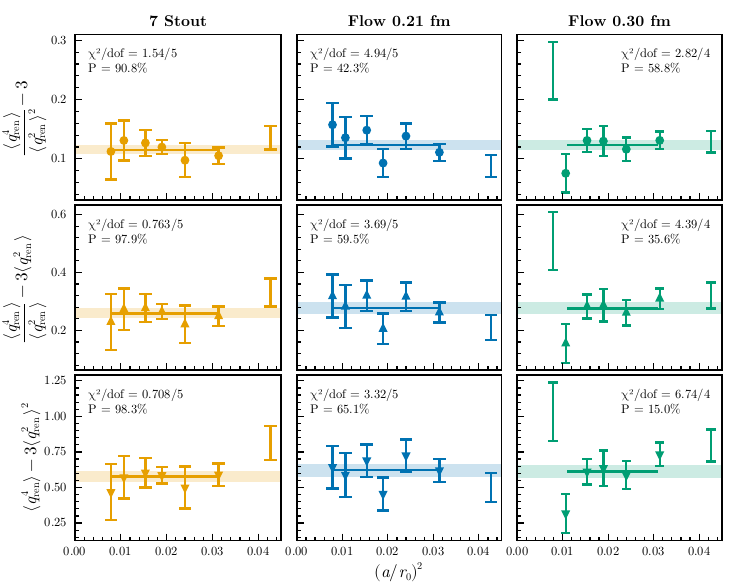}%
\vspace*{-4pt}
\caption{\sl
Continuum extrapolation of three varieties of the excess kurtosis (top, middle, bottom)
for our three smearing strategies (``7\,stout'' left, ``flow\,0.21\,fm'' middle, ``flow\,0.30\,fm'' right).
\label{fig:kurtosis_pub}}
\end{figure}

The measured distribution of $q_\mr{ren}$ is, for each $\be$, precise enough that we may attempt to determine a fourth-order cumulant (with a subsequent continuum limit).
For a random variable $X$ with mean $\mu$ and variance $\si^2$, the ``standard kurtosis'' is defined as $\<Y^4\>$, where $Y=(X-\mu)/\si$.
If $X$ is Gaussian, then $\<Y^2\>=1$ and $\<Y^4\>=3$.
Therefore one defines the ``excess kurtosis'' of $X$ as $\<Y^4\>/\<Y^2\>^2-3$ or $\<Y^4\>/\<Y^2\>-3\<Y^2\>$ or $\<Y^4\>-3\<Y^2\>^2$; each one of these quantities is zero for a Gaussian $X$.

We measured these four varieties of the excess kurtosis of $q_\mr{ren}$ for all our ensembles, the results are given in Tab.~\ref{tab:kurt_data}.
The ``7\,stout'', ``flow\,0.21\,fm'' and ``flow\,0.30\,fm'' results are given as a comma separated list in each cell.
These results are also displayed in Fig.~\ref{fig:kurtosis_pub}; the data seem to approach a well-defined continuum limit.
Upon dropping the coarsest (and for the ``flow\,0.30\,fm'' strategy also the finest) lattice spacing, we get good fits with a single constant for each smearing strategy.
The continuum limit in the fixed physical volume $V=(2.4783r_0)^4$ is found to be
\bea
\<q_\mr{ren}^4\>/\<q_\mr{ren}^2\>^2-3               &=& 0.1150(76)_\mr{7\,stout}, \quad 0.1232(91)_\mr{flow\,0.21\,fm}, \quad 0.1235(90)_\mr{flow\,0.30\,fm}
\label{final_kurt_variety1}
\\
\<q_\mr{ren}^4\>/\<q_\mr{ren}^2\>-3\<q_\mr{ren}^2\> &=&  0.259(17)_\mr{7\,stout}, \quad  0.278(21)_\mr{flow\,0.21\,fm}, \quad  0.276(20)_\mr{flow\,0.30\,fm}
\label{final_kurt_variety2}
\\
\<q_\mr{ren}^4\>-3\<q_\mr{ren}^2\>^2                &=&  0.578(39)_\mr{7\,stout}, \quad  0.622(46)_\mr{flow\,0.21\,fm}, \quad  0.614(45)_\mr{flow\,0.30\,fm}
\label{final_kurt_variety3}
\eea
and we recall that these figures (or the panels in Fig.~\ref{fig:kurtosis_pub}) are vertically correlated but not horizontally.

Regardless which definition of the excess kurtosis is chosen, it is striking to the eye that the horizontal bands of the three smoothing strategies in Fig.~\ref{fig:kurtosis_pub} agree within errors.
This is consistent with what we saw for the second moment (compare Tab.~\ref{tab:results}).
Note that the statement in Refs.~\cite{Ce:2015qha,Bonanno:2023ple} that the universal cutoff $\mu=(8t)^{-1/2}$ of the gradient-flow smoothing strategy has no visible impact on $\<q^2\>$ applies to $\<q^4\>$, too.
In fact, any moment of the global topological charge distribution (\ref{def_partitionfunction}), and thus the distribution itself,
is supposed to enjoy a negligible systematic bias, provided $(8t)^{-1/2}\ll4\pi\Fpi$ holds true \cite{Luscher:2010iy,Luscher:2011bx}.
Our flow-scale choices $\mu=(0.21\fm)^{-1}\simeq940\MeV$ and $\mu=(0.30\fm)^{-1}\simeq660\MeV$ seem to satisfy this criterion.

Last but not least we remind the reader that the final results of both this section and the previous one are subject to potential finite-volume artefacts, which we shall address next.

%%%%%%%%%%%%%%%%%%%%%%%%%%%%%%%%%%%%%%%%%%%%%%%%%%%%%%%%%%%%%%%%%%%%%%%%%%%%%%%

\section{Infinite volume extrapolations \label{sec:infvol}}

%%%%%%%%%%%%%%%%%%%%%%%%%%%%%%%%%%%%%%%%%%%%%%%%%%%%%%%%%%%%%%%%%%%%%%%%%%%%%%%

To get rid of potential finite volume effects in the final results of Secs.~\ref{sec:susc}, \ref{sec:kurt} we need data with different box sizes $L$.
In a gapped theory these finite-volume effects scale asymptotically with $L$ like \cite{Luscher:1985dn}
\beq
\chi_\mr{top}(L)=\chi_\mr{top}(\infty)\cdot\Big[1+\mr{const}\cdot e^{-M_\mr{G}L}\Big]
\label{asymptotic_luscher}
\eeq
where $M_\mr{G}$ is, in our case, the mass of the lightest glueball that couples to $q(x)$.

\begin{table}[tb]
\centering
\begin{tabular}{|cc|ccc|}
\toprule
$L/a$ & $\beta$ & $n_\mr{meas}[n_\mr{sepa}]$ & $\tau_\mr{int}(q_\mr{ren}^2)$ & $Z_q$ (7\,stout) \\
\midrule
12 & 6.1912 & 200000[50] & 1.117(17) & 1.18218(64) \\
14 & 6.1912 & 200000[50] & 1.133(18) & 1.17604(35) \\
16 & 6.1912 & 212789[50] & 1.036(19) & 1.17378(52) \\
18 & 6.1912 & 310191[81] & 0.730(20) & 1.17499(62) \\
20 & 6.1912 & 204997[50] & 1.010(27) & 1.17282(39) \\
24 & 6.1912 & 324894[50] & 1.045(37) & 1.17292(58) \\
28 & 6.1912 & 246512[50] & 1.056(51) & 1.1717(16)  \\
\bottomrule
\end{tabular}
\caption{\sl Details of the ensembles used in the infinite-volume extrapolation.
The format is the same as in Tabs.~\ref{tab:sim_data} and \ref{tab:Z_factor}, except that this time we restrict ourselves to the ``7\,stout'' strategy.
\label{tab:vol_data}}
\end{table}

We select our intermediate coupling ($\be=6.1912$, $18^4$ lattice) in the continuum series, and augment it with smaller/larger boxes as indicated in Tab.~\ref{tab:vol_data}.
We do fewer updates between adjacent measurements [to establish $\ta_\mr{int}(q_\mr{ren}^2)\simeq1$] and (over)compensate this by a larger number of measurements.
Since potential finite volume effects originate in the IR physics of the original (unsmeared) gauge configuration, it is sufficient to do this for one $\be$ and one smoothing strategy (we select ``7\,stout'').
We expect that the factor $Z_q$ in the definition (\ref{def_qren}) of $q_\mr{ren}$ is essentially independent of $L/r_0$, unless the box volume is so small that deconfinement effects are present.
Indeed, the last column of Tab.~\ref{tab:vol_data} confirms this expectation, but the errors grow with the volume.
Therefore we follow Ref.~\cite{Durr:2006ky} and use, in the analysis below, the value $Z_q=1.17499(62)$ of the $18^4$ box for all volumes.

\begin{table}[tb]
\centering
\begin{tabular}{|c@{\hskip 2pt}c|ccccc|}
\toprule
$L/a$ & $\beta$ & $\<q_\mr{ren}^2\>$ & $\<q_\mr{ren}^4\>$ & $\<q_\mr{ren}^4\>/\<q_\mr{ren}^2\>^2-3$ & $\<q_\mr{ren}^4\>/\<q_\mr{ren}^2\>-3\<q_\mr{ren}^2\>$ & $\<q_\mr{ren}^4\>-3\<q_\mr{ren}^2\>^2$ \\
\midrule
12 & 6.1912 & 0.1911(17) & 0.2923(49) & 4.998(79) & 0.956(14) & 0.1827(34) \\
14 & 6.1912 & 0.6689(38) & 1.744(22)  & 0.897(26) & 0.600(18) & 0.401(12)  \\
16 & 6.1912 & 1.3482(63) & 5.884(61)  & 0.238(15) & 0.320(20) & 0.432(27)  \\
18 & 6.1912 & 2.2154(81) & 15.31(13)  & 0.120(12) & 0.265(26) & 0.586(57)  \\
20 & 6.1912 & 3.394(15)  & 35.17(35)  & 0.053(13) & 0.181(43) & 0.61(14)   \\
24 & 6.1912 & 7.019(25)  & 150.0(1.2) & 0.045(11) & 0.318(74) & 2.24(52)   \\
28 & 6.1912 & 13.050(53) & 511.6(4.5) & 0.005(11) & 0.06(14)  & 0.8(1.8)   \\
\bottomrule
\end{tabular}
\caption{\sl
Ensemble average and statistical error of the squared topological charge and the three kurtosis varieties for the ensembles listed in Tab.~\ref{tab:vol_data}.
Throughout the ``7\,stout'' strategy is used.
\label{tab:susc_vol_data}}
\end{table}

With these preparatory steps taken, we determine the second and fourth moments%
\footnote{We checked that the first and third moments $\<q_\mr{ren}\>$ and $\<q_\mr{ren}^3\>$ are zero within errors.}
of the $q_\mr{ren}$ distribution on the ensembles mentioned.
For the latter moment we measure the same three varieties of the excess kurtosis that were studied in Sec.~\ref{sec:kurt}.
The results are listed in Tab.~\ref{tab:susc_vol_data}.

\begin{figure}[tb]
\includegraphics[width=1.0\textwidth]{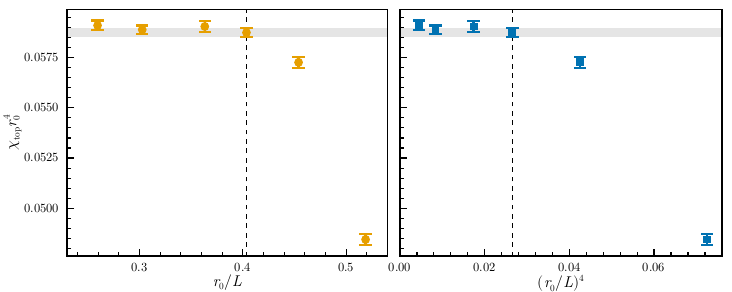}%
\vspace*{-4pt}
\caption{\sl Volume scaling of the topological susceptibility plotted versus $1/L$ (left) and $1/V$ (right).
The standard inverse size/volume used in Sec.~\ref{sec:susc} is marked with a dashed vertical line.
The runs of Ref.~\cite{Durr:2006ky} were performed with a box size $\sim10\%$ smaller, which coincides roughly with the fifth data point.
Our seventh data point is out of scale (both horizontally and vertically).
\label{fig:chir04_vol}}
\end{figure}

\begin{figure}[tb]
\centering
\includegraphics[width=0.9\textwidth]{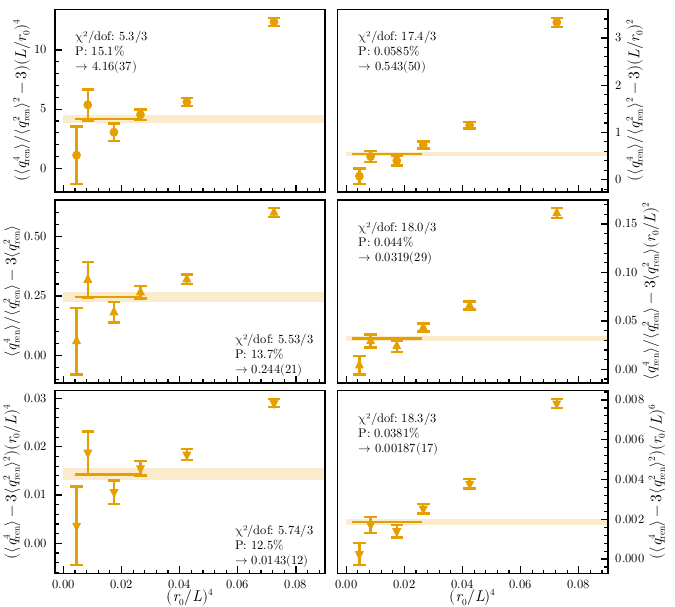}%
\vspace*{-4pt}
\caption{\sl Large volume behavior of the kurtosis varieties (``7\,stout'' smoothing).
In the left panel one variety is multiplied with $(L/r_0)^4$, one unchanged, one divided by $(L/r_0)^4$ (top to bottom).
In the right panel the factors are $(L/r_0)^2$, $(r_0/L)^2$ and $(r_0/L)^6$.
The seventh data point is out of scale.
\label{fig:kurt_vol_V_const_3_rev}}
\end{figure}

%%% 7.263*[0.903,0.038]=[6.558489,0.275994]
%%% [cen,err]=error_div(197.327*6.56,197.327*0.28,0.4757,0.0064)
%%% [cen,err]=error_sub(6.56,0.28,6.12,0.06)

%%% [cen,err]=error_mul(0.4769,[0.0014,0.0010,0.0003],1.0012,0.0002) yields 0.4775(14)(10.488->11) where sqrt(10^2+03^2+01^2)

The first data column of this table is easy to interpret.
The quantity $\<q_\mr{ren}^2\>$ grows linearly with the box volume, so $\chi_\mr{top}=\<q_\mr{ren}^2\>/V$ assumes a finite value in the limit $V\to\infty$, in agreement with theory \cite{Leutwyler:1992yt}.
In Fig.~\ref{fig:chir04_vol} we illustrate this by plotting the topological susceptibility versus $1/L$ and $1/V$.
The standard volume used in Sec.~\ref{sec:susc} (indicated by a dashed vertical line) is almost large enough to avoid finite size effects (within the statistical precision that we have).
Still, fitting the ansatz (\ref{asymptotic_luscher}) to our data with $L/a\geq14$ suggests%
\footnote{As a side result, this fit yields $M_\mr{G}a=0.903(38)$ for the lightest $0^{-+}$ glueball mass. Our result $M_\mr{G}r_0=6.56(28)$
or $M_\mr{G}=2720(120)(40)\MeV$ is compatible with $M_\mr{G}r_0=5.276(45)\times1.160(6)=6.12(6)$ in Ref.~\cite{Athenodorou:2020ani}.}
that there is a factor $1.0048(08)$ between $\chi_\mr{top}(L\simeq2.48r_0)$ and $\chi_\mr{top}(L=\infty)$.
Multiplying the last entry in Tab.~\ref{tab:results} with a quarter root of this factor one finds
\beq
\big[\chi_\mr{top}^{1/4}r_0\big]_\mr{0.30\,fm} = 0.4775(14)_\mr{stat}(11)_\mr{syst} = 0.4775(18)_\mr{tot} %%% note: sqrt(10^2+03^2+01^2)=10.488->11 and sqrt(14^2+10.488^2)=17.493
\label{final_susc_strategy3}
\eeq
for the topological susceptibility with all systematics included (the previous uncertainty and the one from the finite-volume correction factor were added in quadrature).
In physical units we obtain
\beq
\chi_\mr{top}^{1/4}=\frac{0.4775(18)}{0.4757(64)\,\mr{fm}}=198.1(0.7)(2.7)\,\mr{MeV} %%% note: [cen,err]=error_div(197.327*0.4775,197.327*0.0018,0.4757,0.0064)
\label{final_physical}
\eeq
where the first parentheses reflect the total uncertainty of our calculation (with statistics and systematics added in quadrature),
and the second one the uncertainty of $r_0$ from Ref.~\cite{Asmussen:2024hfw}.
The central value and the 0.4\% uncertainty of $\chi_\mr{top}^{1/4} r_0$ in (\ref{final_susc_strategy3}) will be compared to the literature in Sec.~\ref{sec:conc}.

The last three columns of Tab.~\ref{tab:susc_vol_data} are not so easy to interpret.
The $\<q_\mr{ren}^4\>/\<q_\mr{ren}^2\>^2-3$ variety of the excess kurtosis definitely decreases with the box volume,
the $\<q_\mr{ren}^4\>/\<q_\mr{ren}^2\>-3\<q_\mr{ren}^2\>$ variety seems somewhat undecided,
and the $\<q_\mr{ren}^4\>-3\<q_\mr{ren}^2\>^2$ variety likely increases toward the largest volumes.
Hence, we include a factor $(L/r_0)^4$ in the first case, a factor $1$ in the second case, and a factor $(r_0/L)^4$ in the third case to plot them in Fig.~\ref{fig:kurt_vol_V_const_3_rev} (left panels).
It is striking to see that --~with these volume factors included~-- the data for the three varieties resemble each other, except for a change in the vertical scale.
For each variety we get a reasonable fit to a constant, if we include the four largest box volumes, and the resulting value is consistent with other works with similar volumes \cite{Ce:2015qha,Bonati:2015sqt}.

However, from Fig.~\ref{fig:kurt_vol_V_const_3_rev} (left) it is not obvious that the large-volume asymptotic of these quantities is a finite constant.
In App.~\ref{sec:app} we will present an auxiliary study (at coarser lattice spacings) that suggests that one must
include a factor $(L/r_0)^2$, $(r_0/L)^2$ and $(r_0/L)^6$, respectively, in order to obtain a finite value in the infinite-volume limit.
The right panels of Fig.~\ref{fig:kurt_vol_V_const_3_rev} give a preview how this hypothesis works on the present data.
Again, the three excess kurtosis varieties are found to look similar to each other, except for a change in the overall scale.
Each panel includes a fit of the largest four volumes to a constant, which (by the $P$-value) is worse than the respective fit in the left panel.

For now we can only conclude that the data in Tab.~\ref{tab:susc_vol_data} are insufficient to decide on the scaling exponent $\al$ in the asymptotic large-volume behavior
$\propto(L/r_0)^\al$ of the quantities $\<q_\mr{ren}^4\>/\<q_\mr{ren}^2\>^2-3$, $\<q_\mr{ren}^4\>/\<q_\mr{ren}^2\>-3\<q_\mr{ren}^2\>$, and $\<q_\mr{ren}^4\>-3\<q_\mr{ren}^2\>^2$.
This is why we decided to generate another dataset to investigate the situation in more detail (see App.~\ref{sec:app}).
As we shall see, these new data suggest the scaling exponents $\al=-2$, $2$ and $6$ for these excess kurtosis varieties.
In retrospect we will say that the volumes considered in the present section are large enough to reach definite
conclusions for the large-volume behavior of $\<q_\mr{ren}^2\>$, but not for the three excess kurtosis varieties.

%%%%%%%%%%%%%%%%%%%%%%%%%%%%%%%%%%%%%%%%%%%%%%%%%%%%%%%%%%%%%%%%%%%%%%%%%%%%%%%

\section{Discussion \label{sec:discussion}}

%%%%%%%%%%%%%%%%%%%%%%%%%%%%%%%%%%%%%%%%%%%%%%%%%%%%%%%%%%%%%%%%%%%%%%%%%%%%%%%

\begin{table}[tb]
\centering
\begin{tabular}{|c@{\hskip 2pt}c|ccc|}
\toprule
 & & $\chi_\mr{top}^{1/4}r_0$ & $\chi_\mr{top}r_0^4$ & $\chi_\mr{top}^{1/4}\,[\mr{MeV}]$ \\
\midrule
this work                       & (2025) & 0.4775(18) & 0.05199(78) & 198.1(0.7)(2.7) \\
Ref.~\cite{Bonanno:2023ple}     & (2023) & 0.4794(86) & 0.0528(38)  & 198.9(3.6)(2.7) \\
Ref.~\cite{Athenodorou:2021qvs} & (2021) & 0.4926(48) & 0.0589(23)  & 204.3(2.0)(2.8) \\
Ref.~\cite{Athenodorou:2020ani} & (2020) & 0.4857(67) & 0.0557(31)  & 201.5(2.8)(2.7) \\
Ref.~\cite{Bonati:2015sqt}      & (2015) & 0.4708(72) & 0.0491(30)  & 195.3(3.0)(2.7) \\
Ref.~\cite{Ce:2015qha}          & (2015) & 0.4829(40) & 0.0544(18)  & 200.3(1.7)(2.7) \\
Ref.~\cite{Cichy:2015jra}       & (2015) & 0.470(14)  & 0.049(6)    & 195.0(5.8)(2.6) \\
Ref.~\cite{Luscher:2010ik}      & (2010) & 0.498(13)  & 0.0615(64)  & 206.6(5.4)(2.8) \\ %%% note: paper gives 196.5(5.1)/197.327*0.5
Ref.~\cite{Durr:2006ky}         & (2006) & 0.4784(21) & 0.05236(94) & 198.4(0.9)(2.7) \\ %%% note: paper gives 0.05236(71)(62)=0.05236(94)
Ref.~\cite{DelDebbio:2004ns}    & (2004) & 0.4928(62) & 0.059(3)    & 204.4(2.6)(2.8) \\
Ref.~\cite{Giusti:2003gf}       & (2003) & 0.4928(104)& 0.059(5)    & 204.4(4.3)(2.8) \\
Ref.~\cite{DelDebbio:2003rn}    & (2003) & 0.4821(67) & 0.054(3)    & 200.0(3.9)(2.7) \\
\bottomrule
\end{tabular}
\caption{\sl
Summary of $\chi_\mr{top}$ values in $SU(3)$ gauge theory with continuum extrapolation, in units of $r_0^{-4}$.
The results $\chi_\mr{top}^{1/4}/\sqrt{\si}=0.4187(53)$~\cite{Athenodorou:2020ani} and $\chi_\mr{top}^{1/4}/\sqrt{\si}=0.4246(36)$~\cite{Athenodorou:2021qvs} are converted via $r_0\sqrt{\si}=1.160(6)$.
The values in the $\chi_\mr{top}^{1/4}r_0$ column are converted to MeV by means of $r_0=0.4757(64)\,\fm$ \cite{Asmussen:2024hfw}.
\label{tab:comparison}}
\end{table}

%%% cen=[0.4775,0.4794,0.4926,0.4857,0.4708,0.4829,0.470,0.498,0.4784,0.4928,0.4928,0.4821];
%%% err=[0.0018,0.0086,0.0048,0.0067,0.0072,0.0040,0.014,0.013,0.0021,0.0062,0.0104,0.0067];
%%% [cen,err]=error_pow(cen,err,4); obj=[cen;err]';
%%% round(1e5*obj(1,:)), round(1e4*obj(2:6,:)), round(1e+3*obj(7,:)), round(1e+4*obj(8,:)), round(1e5*obj(9,:)), round(1e3*obj(10:12,:))

\begin{figure}[tb]
\centering
\includegraphics[width=0.9\textwidth]{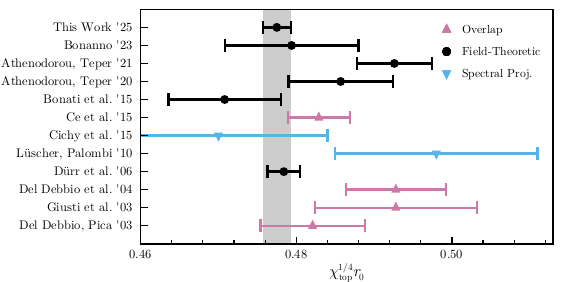}%
\vspace*{-4pt}
\caption{\sl
Graphical display of the results for $\chi_\mr{top}^{1/4}r_0$ as listed in the first column of Tab.~\ref{tab:comparison}.
\label{fig:chi_summary}}
\end{figure}

Our final result for $\chi_\mr{top}^{1/4}r_0$ was given in Eqs.~(\ref{final_susc_strategy3},\ref{final_physical}).
In Tab.~\ref{tab:comparison} this quantity and its fourth power are compared to results in the literature, under the proviso that they come with a continuum extrapolation.
Except for Refs.~\cite{Athenodorou:2020ani,Athenodorou:2021qvs} all these papers effectively compute $\chi_\mr{top}^{1/4}r_0$ or $\chi_\mr{top}r_0^4$.
This is why we ignore their final quote in physical units (if given) and convert, in the last column, all results with the same factor to MeV units.
This factor $r_0^{-1}=414.8(5.6)\MeV$ is taken from Ref.~\cite{Asmussen:2024hfw}; its error bar is reflected by the last parentheses (which dominates the final uncertainty).

%%% [~,~,rel]=error_sub(0.4775,0.0018,[0.4926,0.4928]                    ,[0.0048,0.0062]                    ); 1./rel %%% Athenodorou:2021qvs,DelDebbio:2004ns
%%% [~,~,rel]=error_sub(0.4775,0.0018,[0.4857,0.4829,0.498,0.4928]       ,[0.0067,0.0040,0.013,0.0104]       ); 1./rel %%% Athenodorou:2020ani,Ce:2015qha,Luscher:2010ik,Giusti:2003gf
%%% [~,~,rel]=error_sub(0.4775,0.0018,[0.4794,0.4708,0.470,0.4784,0.4821],[0.0086,0.0072,0.014,0.0021,0.0067]); 1./rel %%% Bonanno:2023ple,Bonati:2015sqt,Durr:2006ky,DelDebbio:2003rn

Considering the $\chi_\mr{top}^{1/4}r_0$ column,
there is some tension ($2.9$ and $2.4$ combined standard deviations)  with Refs.~\cite{Athenodorou:2021qvs},\cite{DelDebbio:2004ns},
reasonable agreement  ($1.2$, $1.2$, $1.6$ and $1.4$ combined sigmas) with Refs.~\cite{Athenodorou:2020ani},\cite{Ce:2015qha}, \cite{Luscher:2010ik},\cite{Giusti:2003gf},
and perfect agreement ($0.2$, $0.9$, $0.5$, $0.3$ and $0.7$ sigmas)   with Refs.~\cite{Bonanno:2023ple},\cite{Bonati:2015sqt},\cite{Cichy:2015jra},\cite{Durr:2006ky},\cite{DelDebbio:2003rn}.
This is what one expects if every paper assesses its uncertainty correctly; see Fig.~\ref{fig:chi_summary} for an overview.
We think that we follow this tradition; we have seven lattice spacings available, down to $a\simeq0.042\fm$, and we use a large variety of fit functions and cuts to determine our systematic uncertainty.

Regarding the excess kurtosis, the situation is different.
The continuum limit in a fixed volume is benevolent (see Fig.~\ref{fig:kurtosis_pub}), but the large-volume scaling behavior is not.
We agree with Refs.~\cite{Ce:2015qha,Bonati:2015sqt} on the value in a volume $V\simeq(2.5r_0)^4$, but we cannot confirm their statement that $\<q^4\>/\<q^2\>-3\<q^2\>$ has a finite value in the limit $L\to\infty$.
Our data in Sec.~\ref{sec:infvol} are inconclusive on this point, and the auxiliary data in App.~\ref{sec:app} suggest that the latter quantity scales like $L^2$.

%%%%%%%%%%%%%%%%%%%%%%%%%%%%%%%%%%%%%%%%%%%%%%%%%%%%%%%%%%%%%%%%%%%%%%%%%%%%%%%

\section{Conclusions \label{sec:conc}}

%%%%%%%%%%%%%%%%%%%%%%%%%%%%%%%%%%%%%%%%%%%%%%%%%%%%%%%%%%%%%%%%%%%%%%%%%%%%%%%

In many respects the present study is an update to Ref.~\cite{Durr:2006ky} which, for many years, was the most precise determination of $\chi_\mr{top}r_0^4$ in $SU(3)$ pure gauge theory.
The present work increases the precision, but only marginally (the overall uncertainty decreases from 1.8\% to 1.5\%).
Our study uses exactly the same methodology as Refs.~\cite{Hoek:1986nd,Lucini:2001ej,DelDebbio:2002xa,Durr:2006ky,Borsanyi:2015cka,Bonati:2014tqa,Alexandrou:2017hqw,Athenodorou:2020ani,Athenodorou:2021qvs} (and likely many more),
with a smoothed-link gluonic definition of the global topological charge, and a careful continuum extrapolation to get rid of the $O(a^2)$ lattice artefacts.
In addition, an extensive finite-volume scaling study was performed.

The point where we differ from previous investigations is that we compare different smoothing strategies in the definition of the topological charge operator.
Our ``7\,stout'' strategy implements the traditional choice where the smoothing radius $\sqrt{8t}$ is held fixed in lattice units (in our case at $2.59a$).
By contrast, our ``0.21\,fm'' and ``0.30\,fm'' strategies increase, on the way to the continuum, the smoothing radius $\sqrt{8t}$ in lattice units so that it stays fixed in physical units (at the distance indicated by the name).
In line with theoretical arguments \cite{Ce:2015qha,Bonanno:2023ple} our data support the idea that there is a joint continuum limit under the auspices of either strategy.

Evidently, the research reported here is a snapshot in the broader context of lattice gluodynamics.
It would be interesting to follow Refs.~\cite{Lucini:2001ej,DelDebbio:2002xa,Vicari:2008jw,Athenodorou:2021qvs} and vary the number of colors beyond $\Nc=3$.
Also quantities with nonzero virtuality would be interesting, such as $\chi_\mr{top}'(p^2)|_{p^2=0}$ \cite{DelDebbio:2002xa,Vicari:2008jw,Bonati:2015sqt,Bonanno:2023ple}
or the glueball mass extracted from the long distance behavior of $q(x)q(0)$.
On the technical level, perhaps an improved field strength definition as advocated in Ref.~\cite{Bilson-Thompson:2002xlt} might be a useful addition.

\subsection*{Acknowlegments}

Computations per performed on (the remnants of) a small PC cluster at the University of Wuppertal that was originally financed by DFG as part of the SFB-TRR-55 grant.

%%%%%%%%%%%%%%%%%%%%%%%%%%%%%%%%%%%%%%%%%%%%%%%%%%%%%%%%%%%%%%%%%%%%%%%%%%%%%%%

\appendix\section{Large-volume scaling test with another dataset\label{sec:app}}

%%%%%%%%%%%%%%%%%%%%%%%%%%%%%%%%%%%%%%%%%%%%%%%%%%%%%%%%%%%%%%%%%%%%%%%%%%%%%%%

\begin{table}[tb]
\centering
\begin{tabular}{|cc|ccc|}
\toprule
$L/a$ & $\beta$ & $n_\mr{meas}[n_\mr{sepa}]$ & $\tau_\mr{int}(q_\mr{ren}^2)$ & $Z_q$ (7\,stout) \\
\midrule
10 & 5.9421 & 281979[10] & 0.7039(76) & 1.28139(96) \\
12 & 5.9421 & 100000[10] & 0.653(11)  & 1.2757(18)  \\
14 & 5.9421 & 535206[10] & 0.6533(48) & 1.2757(79)  \\
16 & 5.9421 & 846551[10] & 0.6485(67) & 1.279(59)   \\
18 & 5.9421 & 420704[10] & 0.6481(76) & 1.13(32)    \\
20 & 5.9421 & 707643[10] & 0.6465(90) & 1.201(12)   \\
22 & 5.9421 & 809898[10] & 0.6369(93) & 1.27(62)    \\
\bottomrule
\end{tabular}
\caption{\sl Details of the post production ensembles; the format is the same as in Tab.~\ref{tab:vol_data}.
\label{tab:postproduction_data}}
\end{table}

In an attempt to improve on the large volume scaling tests presented in Sec.~\ref{sec:infvol},
we decided to generate another series of lattices which would reach toward larger physical volumes.
For the intermediate coupling $\be=6.1912$, as used in that section, we exhausted our computational resources.
Hence, another attempt must use a lower $\be$ (stronger coupling).

\begin{table}[tb]
\centering
\begin{tabular}{|c@{\hskip 2pt}c|ccccc|}
\toprule
$L/a$ & $\beta$ & $\<q_\mr{ren}^2\>$ & $\<q_\mr{ren}^4\>$ & $\<q_\mr{ren}^4\>/\<q_\mr{ren}^2\>^2-3$ & $\<q_\mr{ren}^4\>/\<q_\mr{ren}^2\>-3\<q_\mr{ren}^2\>$ & $\<q_\mr{ren}^4\>-3\<q_\mr{ren}^2\>^2$ \\
\midrule
10 & 5.9421 & 1.1095(38) & 4.167(34)   & 0.385(14)  & 0.427(15) & 0.474(17) \\
12 & 5.9421 & 2.450(13)  & 18.81(23)   & 0.135(20)  & 0.331(48) & 0.81(12)  \\
14 & 5.9421 & 4.546(10)  & 63.13(32)   & 0.0550(74) & 0.250(34) & 1.14(15)  \\
16 & 5.9421 & 7.671(13)  & 178.06(70)  & 0.0261(55) & 0.201(42) & 1.54(33)  \\
18 & 5.9421 & 12.257(31) & 456.7(2.7)  & 0.0408(84) & 0.50(10)  & 6.1(1.3)  \\
20 & 5.9421 & 18.557(36) & 1041.1(4.5) & 0.0221(61) & 0.41(11)  & 7.6(2.1)  \\
22 & 5.9421 & 27.251(49) & 2244.6(9.1) & 0.0237(57) & 0.65(15)  & 17.5(4.2) \\
\bottomrule
\end{tabular}
\caption{\sl
Ensemble average and statistical error of the squared topological charge and the three kurtosis definitions for the ensembles listed in Tab.~\ref{tab:postproduction_data}.
Throughout the ``7\,stout'' strategy is used.
\label{tab:postproduction_susc}}
\end{table}

We select $\be=5.9421$ for which there is a series of 100\,000 lattices in a $12^4$ box available.
We complement this with $L/a=10,14,16,18,20,22$, thus spanning the range $2.07\leq L/r_0 \leq4.54$,
as opposed to $1.65\leq L/r_0 \leq3.86$ in Sec.~\ref{sec:infvol}.
The details of these ``postproduction ensembles'' are listed in Tab.~\ref{tab:postproduction_data}.
A technical point is that the statistical precision of $Z_q$, as listed in this table, degrades for large volumes.
In line with Ref.~\cite{Durr:2006ky} and our procedure in Sec.~\ref{sec:infvol} we use, in the analysis below, the value $Z_q=1.2757(18)$ of the $12^4$ box for all volumes.

\begin{figure}[tb]
\centering
%\includegraphics[width=0.5\textwidth]{vol_scale_r0divL_q2_compare.pdf}%
%\includegraphics[width=0.5\textwidth]{vol_scale_r0divL_kurt1_compare.pdf}\\
%\includegraphics[width=0.5\textwidth]{vol_scale_r0divL_kurt2_compare.pdf}%
%\includegraphics[width=0.5\textwidth]{vol_scale_r0divL_kurt3_compare.pdf}%
%%%
\includegraphics[width=1.0\textwidth]{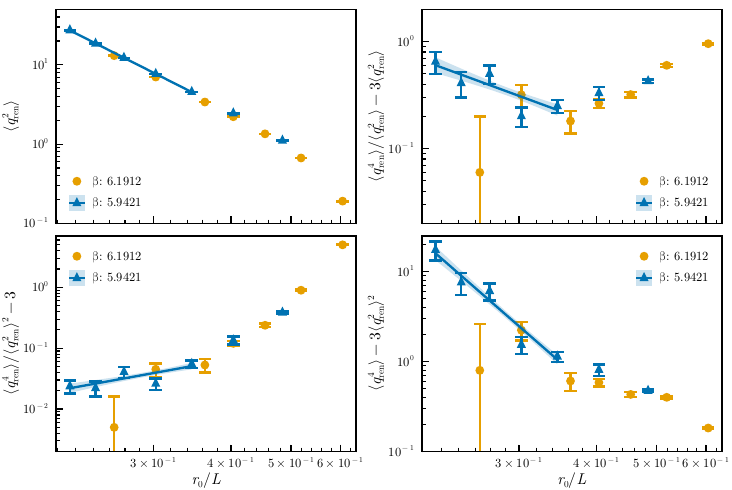}%
\vspace*{-4pt}
\caption{\sl Large volume scaling of the topological susceptibility (top left) and of the three kurtosis varieties (remaining panels) with the ``7\,stout'' strategy.
The new data ($\be=5.9421$, blue triangles) show mild cutoff effects relative to the old ones ($\be=6.1912$, orange circles), but they extend to larger box sizes.
The power-law fits based on the five largest volumes suggest the scaling laws (\ref{largevol_kurtosis}).
\label{fig:large_volume_appendix}}
\end{figure}

On these ensembles we measure the topological susceptibility $\<q_\mr{ren}^2\>/V$ and the same excess kurtosis varieties as in Sec.~\ref{sec:infvol}.
Again we limit ourselves to the ``7\,stout'' smoothing strategy.
The results are listed in Tab.~\ref{tab:postproduction_susc} and shown in Fig.~\ref{fig:large_volume_appendix}.
For comparison the old data ($\be=6.1912$) are included.
For the topological susceptibility some mild discretization effects are visible,
while for the fourth-order cumulants they seem to be negligible at the current level of statistical precision.

% Fits mit 5 Werten: & $chi^2/dof$ & $P$ [\%] & $\alpha$      & $c$         \\
% $<q^2>$            & 11.8/3      & 0.81     & -3.9623(58)   & 0.06753(52) \\
% $<k_1>$            & 4.27/3      & 23.4     & 1.90(53)      & 0.40(26)    \\
% $<k_2>$            & 4.43/3      & 21.9     & -2.07(52)     & 0.027(17)   \\
% $<k_3>$            & 4.28/3      & 23.3     & -5.98(52)     & 0.0019(12)
% %%%
% Fits mit 4 Werten: & $chi^2/dof$ & $P$ [\%] & $\alpha$    \\
% $<q^2>$            & 3.54/2      & 17.0     & -3.9835(94) \\
% $<k_1>$            & 3.88/2      & 14.4     & 1.21(1.2)   \\
% $<k_2>$            & 4.04/2      & 13.3     & -2.74(1.2)  \\
% $<k_3>$            & 3.76/2      & 15.3     & -6.78(1.2)
% %%%
% Fits mit 3 Werten: & $chi^2/dof$ & $P$ [\%] & $\alpha$    \\
% $<q^2>$            & 2.07/1      & 15.0     & -3.998(15)  \\
% $<k_1>$            & 1.06/1      & 30.3     & 2.87(1.6)   \\
% $<k_2>$            & 1.18/1      & 27.7     & -1.15(1.5)  \\
% $<k_3>$            & 1.09/1      & 29.6     & -5.11(1.6)

%Und die neuen Parameter fuer die infinite-volume scaling fits (Plots im Anhang) sind:
%\alpha = 1.85(53), c = 0.36(23) - kurtosis 1
%\alpha = -2.13(52), c = 0.024(15) - kurtosis 2
%\alpha = -6.02(52), c = 0.0018(11) - kurtosis 3

We apply power-law fits $c\,(L/r_0)^\alpha$ to the observables $\<q_\mr{ren}^4\>/\<q_\mr{ren}^2\>^2-3$,
$\<q_\mr{ren}^4\>/\<q_\mr{ren}^2\>-3\<q_\mr{ren}^2\>$ and $\<q_\mr{ren}^4\>-3\<q_\mr{ren}^2\>^2$.
With the five largest volumes included, we find $\al=-1.85(53)$, $2.13(52)$ and $6.02(52)$, respectively, with $P\simeq0.1$ (in all three cases).
And the prefactors $c$ are $0.36(23)$, $0.024(15)$ and $0.0018(11)$, respectively, for these excess kurtosis varieties.

Note that the fitted power $\al$ of the first excess kurtosis variety is negative by almost $4\si$.
Hence, if $\<q_\mr{ren}^4\>/\<q_\mr{ren}^2\>^2-3$ is used to quantify the deviation from a normal distribution,
our fit suggests that any non-Gaussian shape of the topological charge histogram is a finite-volume effect.

In addition, it is worth pointing out that the fitted powers $\al$ are deceptively close%
\footnote{These findings are hard to reconcile with the standard view that $\<q^4\>/\<q^2\>-3\<q^2\>$ tends to a constant value in the $V\to\infty$ limit
\cite{DelDebbio:2002xa,Vicari:2008jw,Ce:2015qha,Bonati:2015sqt,Bonanno:2025wcv}.
Interestingly, our value (\ref{final_kurt_variety2}) is in good agreement with the values $0.233(45),0.259(19)$ obtained in Refs.~\cite{Ce:2015qha,Bonati:2015sqt} in a similar volume $V\simeq(2.5r_0)^4$.
To the best of our knowledge, the volumes considered in this appendix are larger than those of other works, but there is no guarantee that we see the true asymptotic behavior.}
to the integer values $-2$, $+2$ and $+6$, respectively.
In view of the well-known volume scaling law \cite{Leutwyler:1992yt}
\beq
\<q_\mr{ren}^2\> \propto L^4, \quad
\label{largevol_topsusc}
\eeq
for the second cumulant, the fits suggest the scaling laws
\beq
\<q_\mr{ren}^4\>/\<q_\mr{ren}^2\>^2-3 \propto L^{-2}, \quad
\<q_\mr{ren}^4\>/\<q_\mr{ren}^2\>-3\<q_\mr{ren}^2\> \propto L^{2}, \quad
\<q_\mr{ren}^4\>-3\<q_\mr{ren}^2\>^2 \propto L^{6}
\label{largevol_kurtosis}
\eeq
for the three excess kurtosis varieties.
Hence the first relation stipulates that $\<q_\mr{ren}^4\>/\<q_\mr{ren}^2\>^2$ tends, in the infinite-volume limit,
to the value $3$ in such a way that the difference is asymptotically suppressed by two powers of $L$.
Similarly, the second relation says that $\<q_\mr{ren}^4\>/\<q_\mr{ren}^2\>$ and $3\<q_\mr{ren}^2\>$ both grow like
$L^4$, but the difference grows only like $L^2$.
And the third relation suggests that $\<q_\mr{ren}^4\>$ and $3\<q_\mr{ren}^2\>^2$ individually grow like $L^8$,
while the difference stays behind by two powers of $L$.

In the event the conjectures (\ref{largevol_kurtosis}) turn out to be correct, the quantities
\beq
\Big[\<q_\mr{ren}^4\>/\<q_\mr{ren}^2\>^2-3\Big]\frac{L^2}{r_0^2}, \quad
\Big[\<q_\mr{ren}^4\>/\<q_\mr{ren}^2\>-3\<q_\mr{ren}^2\>\Big]\frac{r_0^2}{L^2}, \quad
\Big[\<q_\mr{ren}^4\>-3\<q_\mr{ren}^2\>^2\Big]\frac{r_0^6}{L^6}
\label{largevol_suggestions}
\eeq
assume universal (finite) values in the combined $a\to0,L\to\infty$ limit.
Obviously, in these equations $r_0$ \cite{Sommer:1993ce} may be replaced by another suitable distance,
e.g.\ $t_0^{1/2}$ \cite{Luscher:2010iy,Luscher:2011bx} or $w_0$ \cite{BMW:2012hcm}.
Our results for the respective constants $c$ were mentioned in the text ahead of Eq.~(\ref{largevol_topsusc}).

%%%%%%%%%%%%%%%%%%%%%%%%%%%%%%%%%%%%%%%%%%%%%%%%%%%%%%%%%%%%%%%%%%%%%%%%%%%%%%%

%%%%%%%%%%%%%%%%%%%%%%%%%%%%%%%%%%%%%%%%%%%%%%%%%%%%%%%%%%%%%%%%%%%%%%%%%%%%%%%

\end{document}